%% file: ita16.tex
\documentclass[10pt]{IEEEtran}
\usepackage{graphics, subfigure, color, epsfig, psfrag, amsmath, amsfonts, amssymb, eucal, balance}
\newtheorem{lem}{Lemma}
\newtheorem{prop}{Proposition}
\newtheorem{defn}{Definition}
\newtheorem{rem}{Remark}

\input{rv_defs1}

\begin{document}

\title{A Remark on Channels with Transceiver Distortion}
\author{Wenyi Zhang
\thanks{The author is with Key Laboratory of Wireless-Optical Communications, Chinese Academy of Sciences, and Department of Electronic Engineering and Information Science, University of Science and Technology of China, Hefei 230027, China. Email: {\tt wenyizha@ustc.edu.cn}. The work was supported by National Natural Science Foundation of China through grant 61379003.}
}

\maketitle

\begin{abstract}
Information transmission over channels with transceiver distortion is investigated via generalized mutual information (GMI) under Gaussian input distribution and nearest-neighbor decoding. A canonical transceiver structure in which the channel output is processed by a minimum mean-squared error estimator before decoding is established to maximize the GMI, and the well-known Bussgang's decomposition is shown to be a heuristic that is consistent with the GMI under linear output processing.
\end{abstract}

\section{Introduction}
\label{sec:intro}

A common phenomenon in information transmission over a channel is that the transmitter and the receiver undergo various forms of distortion, which are usually nonlinear, for example, quantization, clipping, saturation, I/Q imbalances, phase oscillation, and so on. A simple and popular approach for handling such channels is linearization, namely, treating the channel output as the linear superposition of the channel input with appropriate scaling and a disturbance. The idea of linearization originates from a well-known result, originally identified by Bussgang \cite{bussgang52:rle} and later recognized as a special case of Price's theorem \cite{price58:it} \cite{rowe82:bstj}, which, for a (continuous-time) stationary Gaussian input process $x(t)$ and a memoryless nonlinearity $h(\cdot)$ such that the output process is $y(t) =  h(x(t))$, indicates that the cross-correlation function between $x(t)$ and $y(t)$ is simply a scaled version of the autocorrelation function $R_{xx}(\tau)$ of $x(t)$, i.e.,
\begin{eqnarray}
\label{eqn:bussgang}
R_{xy}(\tau) = \frac{R_{xy}(0)}{R_{xx}(0)} R_{xx}(\tau).
\end{eqnarray}
A direct consequence of (\ref{eqn:bussgang}) is that the output process $y(t)$ may be linearized as
\begin{eqnarray}
\label{eqn:linearization-bussgang}
y(t) = \frac{R_{xy}(0)}{R_{xx}(0)} x(t) + w(t),
\end{eqnarray}
such that the disturbance process $w(t)$ is uncorrelated with the input process $x(t)$.

When considering information transmission over a channel, the channel output is no longer a deterministic function of the channel input as described by a memoryless nonlinearity. Nevertheless, the basic idea of the linearization in (\ref{eqn:linearization-bussgang}) has been extensively exploited. For example, the clipping process in OFDM systems is directly linearized following (\ref{eqn:linearization-bussgang}) in, e.g., \cite{ochiai02:tcom}; the residual quantization error due to analog-to-digital conversion (ADC) is linearized following (\ref{eqn:linearization-bussgang}) in, e.g., \cite{orhan15:ita}; furthermore, a general linearized model for modeling the composite effect of various forms of transceiver distortion is adopted in \cite{bjornson14:it}, wherein the disturbance is assumed to be not only uncorrelated with, but also independent of, the channel input.

In this paper, we address the following questions. First, is there an information-theoretic interpretation of the Bussgang's decomposition like (\ref{eqn:linearization-bussgang})? Second, is there any decomposition that improves upon (\ref{eqn:linearization-bussgang})? Our approach is based on an analysis of the generalized mutual information (GMI), which is an achievable rate of information transmission under mismatched decoding metrics, i.e., mismatched decoding (see, e.g., \cite{lapidoth98:it} and references therein).

\section{Memoryless Distortion}
\label{sec:memoryless}

\subsection{Preliminary}
\label{subsec:preliminary}

In this subsection, we briefly review the main result of \cite{zhang12:tcom}. Consider a discrete-time channel whose real-valued input sequence is $x_k$, $k = 1, 2, \ldots$, and each input $x_k$ undergoes a memoryless stochastic transformation to yield the corresponding real-valued output $y_k$. The transmission block length is $n$ and the information rate is $R$ so that there are $2^{nR}$ messages. The codeword for each message is drawn uniformly from a Gaussian ensemble with variance $\mathcal{E}_s$, i.e., $\underline{\rvx} = [\rvx_1, \rvx_2, \ldots, \rvx_n] \sim \mathcal{N}(0, \mathcal{E}_s \mathbf{I}_n)$. Upon receiving the channel output sequence $y_k$, $k = 1, 2, \ldots, n$, the decoder is a nearest-neighbor decoder which implements
\begin{eqnarray}
\label{eqn:nn-decoder}
\hat{m} &=& \mathrm{arg}\min_{m \in \{1, \ldots, 2^{nR} \}} D(m),\\
\label{eqn:distance-metric}
D(m) &=& \frac{1}{n} \sum_{k = 1}^n [y_k - a x_k(m)]^2.
\end{eqnarray}
Here, $m$ is the index of the transmitted message, $x_k(m) \sim \mathcal{N}(0, \mathcal{E}_s)$ is the transmitted symbol for the $m$-th codeword at time $k$. Additionally, a parameter $a$ is included for optimizing the transmission rate.

Note that in the transmission system described above, the nearest-neighbor decoder is generally not the maximum-likelihood decoder, i.e., the decoder is mismatched to the channel. For such mismatched decoding problems, determining the maximally achievable information rate is still an open problem, and its achievable lower bounds have been established; see, e.g., \cite{lapidoth98:it} and references therein. The generalized mutual information (GMI) is an achievable information rate, and is indeed the maximally achievable information rate such that the average probability of decoding error asymptotically vanishes as the transmission block length grows without bound, when the codewords are randomly drawn from the specified ensemble; see, e.g., \cite[pp. 1121-1122]{lapidoth02:it}.

A tractable expression of the GMI is obtained in \cite{zhang12:tcom}, as follows.
\begin{prop}
\label{prop:gmi-baseline}
\cite[Prop. 1]{zhang12:tcom} For the transmission system described above, where the channel input $\rvx$ follows an independent and identically distributed (i.i.d.) Gaussian ensemble with mean zero and variance $\mathcal{E}_s$ and the channel output $\rvy$ undergoes a nearest-neighbor decoder as (\ref{eqn:nn-decoder}), the GMI is
\begin{eqnarray}
\label{eqn:gmi-baseline}
I_\mathrm{GMI} &=& \frac{1}{2} \log \left(1 + \frac{\Delta}{1 - \Delta}\right),\\
\label{eqn:Delta}
\Delta &=& \frac{\left\{\mathbf{E}\left[\rvx \rvy\right]\right\}^2}{\mathcal{E}_s \mathbf{E}\left[\rvy^2\right]}.
\end{eqnarray}
\end{prop}

\subsection{Correlation Ratio and Canonical Receiver}

Instead of using the raw channel output $\rvy$, if we process it using a mapping $g$ so as to modify the distance metric in (\ref{eqn:distance-metric}) into
\begin{eqnarray}
D_g(m) = \frac{1}{n} \sum_{k = 1}^n \left[g(y_k) - a x_k(m)\right]^2,
\end{eqnarray}
then as a direct application of Proposition \ref{prop:gmi-baseline} we have the following result.
\begin{prop}
\label{prop:gmi-g}
Under the setting of Proposition \ref{prop:gmi-baseline}, except that the channel output $\rvy$ is further mapped into $g(\rvy)$ before fed into the nearest-neighbor decoder, the GMI is
\begin{eqnarray}
\label{eqn:gmi-g}
I_{\mathrm{GMI}, g} &=& \frac{1}{2} \log \left(1 + \frac{\Delta_g}{1 - \Delta_g}\right),\\
\Delta_g &=& \frac{\left\{\mathbf{E}\left[\rvx g(\rvy)\right]\right\}^2}{\mathcal{E}_s \mathbf{E}\left[g(\rvy)^2\right]}.
\end{eqnarray}
\end{prop}

Hence, a natural problem is to optimize $g$ so as to maximize $I_{\mathrm{GMI}, g}$, and this is equivalent to maximizing $\Delta_g$. Interestingly, the square root of the maximum of $\Delta_g$ is exactly the so-called correlation ratio of $\rvx$ on $\rvw$, a quantity introduced by K. Pearson and further studied by A. R\'enyi \cite{renyi59:amash}. This relationship is detailed in the following.
\begin{defn}
\label{defn:corr-ratio}
\cite[Eqn. (1.7)]{renyi59:amash} For two random variables $\rvu$ and $\rvv$, the correlation ratio $\Theta_\rvv(\rvu)$ of $\rvu$ on $\rvv$ is defined as
\begin{eqnarray}
\Theta_\rvv(\rvu) = \sqrt{\frac{\mathrm{var} \mathbf{E}[\rvu|\rvv]}{\mathrm{var} \rvu}},
\end{eqnarray}
if $\mathrm{var} \rvu$ exists and is strictly positive.
\end{defn}

It is clear that $\Theta_\rvv(\rvu)$ lies between zero and one, taking value one if and only if $\rvu$ is a Borel-measurable function of $\rvv$, and taking value zero if (but not only if) $\rvu$ and $\rvv$ are independent. Furthermore, R\'enyi established the following relationship.
\begin{lem}
\label{lem:renyi-corr-ratio}
\cite[Thm. 1]{renyi59:amash} For two random variables $\rvu$ and $\rvv$, if the mean and variance of $\rvu$ exist, we have
\begin{eqnarray}
\label{eqn:renyi-corr-ratio}
\Theta_\rvv(\rvu) = \sup_{g} \left|\frac{\mathbf{E}[\rvu g(\rvv)] - \mathbf{E}[\rvu] \mathbf{E}[g(\rvv)]}{\sqrt{\mathrm{var} \rvu \mathrm{var} g(\rvv)}}\right|,
\end{eqnarray}
where $g$ runs over all Borel-measurable real functions such that the mean and variance of $g(\rvv)$ exist. The supremum of (\ref{eqn:renyi-corr-ratio}) is attainable, if and only if $g(\rvv) = c \mathbf{E}[\rvu|\rvv] + b$ where $c \neq 0$ and $b$ are arbitrary constants.
\end{lem}

Back to the setting of Proposition \ref{prop:gmi-g}, applying Lemma \ref{lem:renyi-corr-ratio} and Definition \ref{defn:corr-ratio}, we have at once that when $g(\rvy) = \mathbf{E}[\rvx|\rvy]$, $\Delta_g$ is maximized as
\begin{eqnarray}
\max_g \Delta_g = \Theta_\rvy^2 (\rvx) = \frac{\mathrm{var} \mathbf{E}[\rvx|\rvy]}{\mathcal{E}_s}.
\end{eqnarray}
Clearly, $g(\rvy) = \mathbf{E}[\rvx|\rvy]$ is the minimum mean-squared error (MMSE) estimate of $\rvx$ upon observing $\rvy$. Let us thus introduce the following ``canonical decomposition'' of $\rvx$ as
\begin{eqnarray}
\label{eqn:reverse-Bussgang}
\rvx = \mathbf{E}[\rvx|\rvy] + \tilde{\rvx},
\end{eqnarray}
in which the estimation error $\tilde{\rvx}$ is uncorrelated with the MMSE estimate $\mathbf{E}[\rvx|\rvy]$. If we interpret the term $\Delta_g/(1 - \Delta_g)$ inside the logarithm of (\ref{eqn:gmi-g}) as the ``effective signal-to-noise ratio (SNR)'', then the maximally achievable effective SNR is
\begin{eqnarray}
\label{eqn:eff-SNR-MMSE}
\max_g \frac{\Delta_g}{1 - \Delta_g} &=& \frac{\Theta_\rvy^2 (\rvx)}{1 - \Theta_\rvy^2 (\rvx)}\nonumber\\
&=& \frac{\mathrm{var} \mathbf{E}[\rvx|\rvy]}{\mathcal{E}_s - \mathrm{var} \mathbf{E}[\rvx|\rvy]}\nonumber\\
&=& \frac{\mathrm{var} \mathbf{E}[\rvx|\rvy]}{\mathrm{var} \tilde{\rvx}}\nonumber\\
&=& \frac{\mathcal{E}_s - \mathsf{mmse}}{\mathsf{mmse}},
\end{eqnarray}
where we use $\mathsf{mmse}$ to denote the MMSE, $\mathrm{var} \tilde{\rvx}$.

\begin{figure}
\centering
\includegraphics[width=3in]{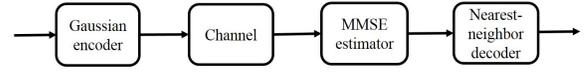}
\caption{Canonical transceiver structure.}
\label{fig:canonical}
\end{figure}

Therefore, we have the following result.
\begin{prop}
\label{prop:canonical}
The maximally achievable effective SNR of the transmission system in Section \ref{subsec:preliminary}, as given by (\ref{eqn:eff-SNR-MMSE}), is simply the ratio between the power of the MMSE estimate and the power of the estimation error (i.e., the MMSE), and is further achieved by the canonical transceiver structure shown in Figure \ref{fig:canonical}.
\end{prop}

\begin{rem}
It is interesting to note that unlike the data processing inequality which asserts that processing the channel output cannot increase the input-output mutual information, for GMI, the preceding analysis reveals that processing the channel output may be beneficial.
\end{rem}

\begin{rem}
For the special case of linear Gaussian channels, $\rvy = \rvx + \rvz$ where $\rvz \sim \mathcal{N}(0, \sigma^2)$ is i.i.d., it can be readily verified that the canonical transceiver structure in Proposition \ref{prop:canonical} leads to $\max_g \Delta_g/(1 - \Delta_g) = \mathcal{E}_s/\sigma^2$, thus restoring the classical additive white Gaussian noise (AWGN) channel capacity result. This is also consistent with the well-known fact that MMSE estimation is information lossless for linear Gaussian channels.
\end{rem}

\begin{rem}
The result obtained here also leads to a special case of the estimation counterpart of Fano's inequality. Noting that the GMI is a lower bound of the mutual information $I(\rvx; \rvy)$ under $\rvx \sim \mathcal{N}(0, \mathcal{E}_s)$, we have
\begin{eqnarray}
\frac{1}{2} \log\left(1 + \frac{\mathcal{E}_s - \mathsf{mmse}}{\mathsf{mmse}} \right) &\leq& I(\rvx; \rvy);\nonumber\\
\mbox{i.e.,}\quad \mathsf{mmse} &\geq& \mathcal{E}_s e^{-2 I(\rvx; \rvy)}\nonumber\\
= \frac{\mathcal{E}_s}{e^{2 h(\rvx)}} e^{2 h(\rvx|\rvy)} &=& \frac{1}{2\pi e} e^{2 h(\rvx|\rvy)},
\end{eqnarray}
which is exactly the conditional estimation counterpart of Fano's inequality \cite[Cor. of Thm. 8.6.6]{cover06:book} specialized to $\rvx \sim \mathcal{N}(0, \mathcal{E}_s)$.
\end{rem}

\subsection{Linear Processing and Bussgang's Decomposition}

In practice, a linear estimator is often employed since computing the nonlinear MMSE estimate is typically complicated and even intractable. For the scalar channel output $\rvy$, when the mapping $g$ is linear (i.e., scaling by a constant coefficient), it is readily verified that the value of $\Delta_g$ in Proposition \ref{prop:gmi-g} is always the same as $\Delta$ in Proposition \ref{prop:gmi-baseline}. In particular, the following result holds.
\begin{prop}
\label{prop:gmi-lmmse}
Under the setting of Proposition \ref{prop:gmi-g}, except that the mapping $g$ is restricted to be a linear scaling of the channel output $\rvy$, the GMI is the same as that in Proposition \ref{prop:gmi-baseline}, and the effective SNR is
\begin{eqnarray}
\label{eqn:eff-SNR-LMMSE}
\frac{\Delta}{1 - \Delta} = \frac{\mathcal{E}_s - \mathsf{lmmse}}{\mathsf{lmmse}},
\end{eqnarray}
where we use $\mathsf{lmmse}$ to denote the mean-squared error of the linear MMSE estimator of $\rvx$ upon observing $\rvy$.
\end{prop}
{\it Proof:} A straightforward calculation shows that
\begin{eqnarray}
\Delta = \frac{1 - \mathsf{lmmse}/\mathcal{E}_s}{\mathsf{lmmse}/\mathcal{E}_s},
\end{eqnarray}
and the proposition readily follows.
$\Box$

\begin{figure}
\centering
\includegraphics[width=3in]{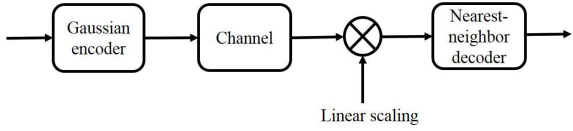}
\caption{Transceiver structure under linear output processing.}
\label{fig:linear}
\end{figure}

Comparing (\ref{eqn:eff-SNR-MMSE}) and (\ref{eqn:eff-SNR-LMMSE}), the loss due to linear processing is revealed, which is exactly due to the loss in replacing the MMSE estimator by the linear MMSE estimator. For channels with nonlinear transceiver distortion these two estimators are different and the loss may be noticeable. The relationship (\ref{eqn:eff-SNR-LMMSE}) is clear when we decompose the channel input $\rvx$ as
\begin{eqnarray}
\label{eqn:linear-reverse-Bussgang}
\rvx = \frac{\mathbf{E}[\rvx \rvy]}{\mathbf{E}[\rvy^2]} \rvy + \tilde{\rvx},
\end{eqnarray}
i.e., the sum of the linear MMSE estimate of $\rvx$ and the estimation error. The effective SNR expression (\ref{eqn:eff-SNR-LMMSE}) is thus the ratio between the power of the linear MMSE estimate and the power of the estimation error, i.e., the LMMSE. The corresponding transceiver structure is illustrated in Figure \ref{fig:linear}.

\begin{rem}
Now we address the questions regarding the Bussgang's decomposition introduced in Section \ref{sec:intro}. The Bussgang's decomposition for an input-output relationship $\rvx \rightarrow \rvy$ can be written as
\begin{eqnarray}
\label{eqn:bussgang-decomp}
\rvy = \frac{\mathbf{E}[\rvx \rvy]}{\mathcal{E}_s} \rvx + \rvw,
\end{eqnarray}
so that the residual $\rvw$ is uncorrelated with $\rvx$. Instead, both (\ref{eqn:reverse-Bussgang}) and (\ref{eqn:linear-reverse-Bussgang}) decompose the channel input $\rvx$, rather than the channel output $\rvy$. Nevertheless, if we view (\ref{eqn:bussgang-decomp}) as an additive noise channel and adopt the nearest-neighbor decoder (\ref{eqn:nn-decoder}) with $a = \mathbf{E}[\rvx \rvy]/\mathcal{E}_s$, i.e., the ``channel coefficient'' in (\ref{eqn:bussgang-decomp}), then from \cite[Prop. 1]{zhang12:tcom}, this choice of $a$ exactly achieves the performance in Proposition \ref{prop:gmi-baseline}, i.e., (\ref{eqn:eff-SNR-LMMSE}).

So for the questions in Section \ref{sec:intro}, we have:
\begin{itemize}
\item The Bussgang's decomposition does have an information-theoretic interpretation, because under i.i.d. Gaussian input, a nearest-neighbor decoder viewing the decomposed channel model as an additive noise channel achieves the GMI (\ref{eqn:gmi-baseline}) with effective SNR (\ref{eqn:eff-SNR-LMMSE}).
\item It is possible to improve upon the Bussgang's decomposition, by following the canonical transceiver structure in Figure \ref{fig:canonical}, which includes an MMSE estimator between the channel output and the nearest-neighbor decoder, and the improved performance is described in Proposition \ref{prop:canonical}.
\end{itemize}
\end{rem}

\section{Distortion with Memory}
\label{sec:memory}

The analysis in Section \ref{sec:memoryless} can be extended to the more general case where the transceiver distortion has memory, for modeling transceivers whose responses are time-varying. Consider a discrete-time channel whose real-valued input/output sequence is $x_k$/$y_k$, $k = 1, 2, \ldots$. The setup is similar to that in Section \ref{sec:memoryless}, except that here the i.i.d. Gaussian input $\{\rvx_k\}$ leads to a stationary and ergodic output process $\{\rvy_k\}$. The decoder is a modified nearest-neighbor decoder which implements
\begin{eqnarray}
\label{eqn:nn-decoder-memory}
\hat{m} &=& \mathrm{arg}\min_{m \in \{1, \ldots, 2^{nR} \}} D_g(m),\\
\label{eqn:distance-metric-memory}
D_g(m) &=& \frac{1}{n} \sum_{k = 1}^n \|g(\underline{y}_k) - a \underline{x}_k(m)\|^2.
\end{eqnarray}
Here, the idea of exploiting the channel memory is to process the channel input/output sequences in segments, so that $\underline{x}_k(m)$ and $\underline{y}_k$ are of length $L$. The mapping $g$ maps the length-$L$ $\underline{y}_k$ into another length-$L$ vector $g(\underline{y}_k)$. Note that the modified nearest-neighbor decoder (\ref{eqn:nn-decoder-memory}) views the channel uses as length-$L$ ``super-symbols'', and thus the resulting GMI needs to be scaled by $L$. We will investigate the performance with $g$ optimized and as $L \rightarrow \infty$ and $n \rightarrow \infty$.
\begin{prop}
Consider the transmission system described above, where the channel input follows an i.i.d. Gaussian ensemble with mean zero and variance $\mathcal{E}_s$ and the channel output process $\{\rvy_k\}$ undergoes a modified nearest-neighbor decoder as (\ref{eqn:nn-decoder-memory}). Assume that the normalized MMSE of estimating $\underline{\rvx}$ upon observing $\underline{\rvy}$ has a limit as $L \rightarrow \infty$, i.e.,
\begin{eqnarray}
\mathrm{mmse} = \lim_{L \rightarrow \infty} (1/L) \mathbf{E}[\left\|\underline{\rvx} - \mathbf{E}[\underline{\rvx}|\underline{\rvy}]\right\|^2].
\end{eqnarray}
The GMI optimized over $g$ as $L \rightarrow \infty$ is
\begin{eqnarray}
I_\mathrm{GMI} = \frac{1}{2} \log\left(1 + \frac{\mathcal{E}_s - \mathrm{mmse}}{\mathrm{mmse}}\right).
\end{eqnarray}
\end{prop}
{\it Proof:} The proof essentially follows the same line as \cite[Thm. 3.0.1]{lapidoth02:it} and \cite[Prop. 1]{zhang12:tcom}. Fix $L$, $g$ and $a$. Without loss of generality, assume that $m = 1$ is the transmitted message. So the distance metric with $m = 1$ satisfies
\begin{eqnarray}
\lim_{n \rightarrow \infty} D_g(1) = \mathbf{E}\left[\|g(\underline{\rvy}) - a \underline{\rvx}\|^2\right], \quad \mbox{a.s.}
\end{eqnarray}
The GMI is then given by
\begin{eqnarray}
I_{\mathrm{GMI}, L, g, a} &=& \sup_{\theta < 0} \left\{\theta \mathbf{E}\left[\|g(\underline{\rvy}) - a \underline{\rvx}\|^2\right] - \Lambda(\theta)\right\},\\
\Lambda(\theta) &=& \lim_{n \rightarrow \infty} \frac{1}{n} \Lambda_n(n\theta),\\
\Lambda_n(n\theta) &=& \log \mathbf{E}\left[e^{n\theta D_g(m)} | \{\underline{\rvy}_k\}\right], \;\; \forall m \neq 1.\label{eqn:Lambda_n_memory}
\end{eqnarray}
The expectation in (\ref{eqn:Lambda_n_memory}) can be evaluated following
\begin{eqnarray}
&&\mathbf{E}\left[e^{n\theta D_g(m)} | \{\underline{\rvy}_k\}\right] = \prod_{k = 1}^n \mathbf{E}\left[e^{\theta \|g(\underline{\rvy}_k) - a \underline{\rvx}_k(m)\|^2}|\underline{\rvy}_k \right]\nonumber\\
&=& \prod_{k = 1}^n \prod_{l = 1}^L \mathbf{E}\left[e^{\theta \left[g(\underline{\rvy}_k)_l - a \underline{\rvx}_{k, l}(m)\right]^2}|\underline{\rvy}_k \right]\nonumber\\
&=& \prod_{k = 1}^n \prod_{l = 1}^L \frac{1}{\sqrt{1 - 2\theta a^2 \mathcal{E}_s}} \exp\left(\frac{\theta [g(\underline{\rvy}_k)_l]^2}{1 - 2\theta a^2 \mathcal{E}_s}\right)\nonumber\\
&=& (1 - 2\theta a^2 \mathcal{E}_s)^{-nL/2} \exp\left(\sum_{k = 1}^n \sum_{l = 1}^L \frac{\theta [g(\underline{\rvy}_k)_l]^2}{1 - 2\theta a^2 \mathcal{E}_s} \right);\nonumber\\
\end{eqnarray}
that is,
\begin{eqnarray}
\Lambda_n(n\theta) = \frac{\theta \sum_{k = 1}^n \|g(\underline{\rvy}_k)\|^2}{1 - 2\theta a^2 \mathcal{E}_s} - \frac{nL}{2} \log (1 - 2\theta a^2 \mathcal{E}_s),
\end{eqnarray}
leading to
\begin{eqnarray}
\Lambda(\theta) = \frac{\theta \mathbf{E} \left[\|g(\underline{\rvy})\|^2\right]}{1 - 2\theta a^2 \mathcal{E}_s} - \frac{L}{2} \log (1 - 2\theta a^2 \mathcal{E}_s), \;\; \mbox{a.s.}
\end{eqnarray}
So the GMI (after scaling by $L$) is
\begin{eqnarray}
I_{\mathrm{GMI}, L, g, a} = \sup_{\theta < 0} \left\{\frac{1}{2} \log (1 - 2\theta a^2 \mathcal{E}_s) +\right.\nonumber\\
 \left. (\theta/L)\mathbf{E}\left[\|g(\underline{\rvy}) - a \underline{\rvx}\|^2\right]
 - \frac{(\theta/L)\mathbf{E} \left[\|g(\underline{\rvy})\|^2\right]}{1 - 2\theta a^2 \mathcal{E}_s}
\right\}.
\end{eqnarray}
Maximizing $I_{\mathrm{GMI}, L, g, a}$ over $\theta$ and $a$ is the vector extension of the problem solved in \cite[Eqn. (77)-(83)]{zhang12:tcom}, and the solution procedure is essentially identical. The optimal $a$ is $a_\mathrm{opt} = \mathbf{E}[\underline{\rvx}^t g(\underline{\rvy})]/(L \mathcal{E}_s)$, and
\begin{eqnarray}
\label{eqn:gmi-memory-a-only}
\max_{a} I_{\mathrm{GMI}, L, g, a} &=& \frac{1}{2} \log\left(1 + \frac{\Delta_{L, g}}{1 - \Delta_{L, g}}\right),\\
\Delta_{L, g} &=& \frac{\left\{\mathbf{E}[\underline{\rvx}^t g(\underline{\rvy})]\right\}^2}{L \mathcal{E}_s \mathbf{E}[\|g(\underline{\rvy})\|^2]}.
\end{eqnarray}
The proof of Lemma \ref{lem:renyi-corr-ratio} \cite[Thm. 1]{renyi59:amash} applies to the maximization of $\Delta_{L, g}$, leading to
\begin{eqnarray}
\max_g \Delta_{L, g} = \frac{\mathrm{tr}\left[\mathrm{cov} \mathbf{E}[\underline{\rvx}|\underline{\rvy}]\right]}{L \mathcal{E}_s},
\end{eqnarray}
achieved by $g_\mathrm{opt}(\underline{\rvy}) = \mathbf{E}[\underline{\rvx}|\underline{\rvy}]$. The effective SNR in (\ref{eqn:gmi-memory-a-only}) is thus
\begin{eqnarray}
\frac{\Delta_{L, g_\mathrm{opt}}}{1 - \Delta_{L, g_\mathrm{opt}}} = \frac{\mathcal{E}_s - \mathrm{mmse}_L}{\mathrm{mmse}_L},
\end{eqnarray}
where $\mathrm{mmse}_L = (1/L) \mathbf{E}[\left\|\underline{\rvx} - \mathbf{E}[\underline{\rvx}|\underline{\rvy}]\right\|^2]$ is the normalized MMSE of estimating $\underline{\rvx}$ upon observing $\underline{\rvy}$. Letting $L \rightarrow \infty$ hence completes the proof.
$\Box$

\section*{Acknowledgement}

Stimulating discussions with Dongning Guo, H. Vincent Poor, and Cong Shen are gratefully acknowledged.


\end{document}

%% file: rv_defs1.tex
\DeclareMathAlphabet{\eurm}{U}{eur}{m}{n}
\DeclareMathAlphabet{\mathbsf}{OT1}{cmss}{bx}{n}
\DeclareMathAlphabet{\mathssf}{OT1}{cmss}{m}{sl}
\DeclareMathAlphabet{\mathcsf}{OT1}{cmss}{sbc}{n}

\newcommand{\randomvalue}[1]{\eurm{\uppercase{#1}}}

\DeclareSymbolFont{bsfletters}{OT1}{cmss}{bx}{n}  
\DeclareSymbolFont{ssfletters}{OT1}{cmss}{m}{n}
\DeclareMathSymbol{\bsfGamma}{0}{bsfletters}{'000}
\DeclareMathSymbol{\ssfGamma}{0}{ssfletters}{'000}
\DeclareMathSymbol{\bsfDelta}{0}{bsfletters}{'001}
\DeclareMathSymbol{\ssfDelta}{0}{ssfletters}{'001}
\DeclareMathSymbol{\bsfTheta}{0}{bsfletters}{'002}
\DeclareMathSymbol{\ssfTheta}{0}{ssfletters}{'002}
\DeclareMathSymbol{\bsfLambda}{0}{bsfletters}{'003}
\DeclareMathSymbol{\ssfLambda}{0}{ssfletters}{'003}
\DeclareMathSymbol{\bsfXi}{0}{bsfletters}{'004}
\DeclareMathSymbol{\ssfXi}{0}{ssfletters}{'004}
\DeclareMathSymbol{\bsfPi}{0}{bsfletters}{'005}
\DeclareMathSymbol{\ssfPi}{0}{ssfletters}{'005}
\DeclareMathSymbol{\bsfSigma}{0}{bsfletters}{'006}
\DeclareMathSymbol{\ssfSigma}{0}{ssfletters}{'006}
\DeclareMathSymbol{\bsfUpsilon}{0}{bsfletters}{'007}
\DeclareMathSymbol{\ssfUpsilon}{0}{ssfletters}{'007}
\DeclareMathSymbol{\bsfPhi}{0}{bsfletters}{'010}
\DeclareMathSymbol{\ssfPhi}{0}{ssfletters}{'010}
\DeclareMathSymbol{\bsfPsi}{0}{bsfletters}{'011}
\DeclareMathSymbol{\ssfPsi}{0}{ssfletters}{'011}
\DeclareMathSymbol{\bsfOmega}{0}{bsfletters}{'012}
\DeclareMathSymbol{\ssfOmega}{0}{ssfletters}{'012}

\newcommand{\rvu}{{\randomvalue{u}}}	

\newcommand{\rvv}{{\randomvalue{v}}}	

\newcommand{\rvw}{{\randomvalue{w}}}	

\newcommand{\rvx}{{\randomvalue{x}}}	

\newcommand{\rvy}{{\randomvalue{y}}}	

\newcommand{\rvz}{{\randomvalue{z}}}	

